\documentstyle[10pt,epsfig,a4wide]{article}

\title{Varying Constants in Brane World Scenarios}
\author{Ph.~Brax \\
{\it Service de Physique Th\'eorique}\\
{\it CEA-Saclay}\\
{\it F-91191, Gif/Yvette cedex, France}\\
{\it }\\
C.~van~de~Bruck, A.~C.~Davis, C.~S.~Rhodes \\
{\it Department of Applied Mathematics and Theoretical Physics}\\
{\it Centre for Mathematical Sciences}\\
{\it Wilberforce Road, University of Cambridge}\\
{\it CB3 0WA, United Kingdom}}

\begin{document}
\maketitle

\begin{abstract}
Higher--dimensional theories imply that some constants, such as the 
gravitational constant and the strength of the gauge--couplings, are
not fundamental constants. Instead they are related to the sizes of 
the extra--dimensional space, which are moduli fields in the
four--dimensional effective theory. We study the cosmological 
evolution of the moduli fields appearing in brane world scenarios 
and discuss the implications for varying constants. 
\end{abstract}

\section{Introduction}
Theories based on the idea of extra dimensions predict the existence
of scalar fields (also called moduli fields), which 
are related to the size of the higher--dimensional space. Classically
these fields do not have a potential, but they might acquire one due to
supersymmetry breaking or quantum mechanical effects. However, 
a potential does not necessarily guarantee stabilization of the fields. 

From the phenomenological point of view there are motivations to study 
the cosmological evolution of moduli fields. First, it is well known
that in scalar--tensor theories of gravity 
there exists an attractor mechanism which 
drives the theory towards general relativity \cite{Damour}. Thus, it
might well be 
that such attractor solutions exist for moduli fields in higher
dimensional theories as well, which would
drive the theory towards those points in moduli space which 
are consistent with observations. Another motivation is the
claim of a varying fine--structure constant made in \cite{Webb}. 
If this observation is confirmed, this would point towards a theory 
in which some of the moduli fields are not stabilised
but are dynamical at least during some part of the cosmological 
history. 

In this paper we consider a class of brane world models and the 
cosmological evolution of moduli fields. In particular we discuss 
which constants necessarily vary in the brane world context. We 
begin with a description of the effective theory in four dimensions.

\section{The Effective Theory at Low Energies}
We consider a five--dimensional theory with two branes and a bulk 
scalar field. The bulk scalar field induces 
tension on each branes with the same amplitude but with different 
signs ($U_1=-U_2$) (see Figure 1). 

\begin{figure}
\centerline{\includegraphics[width=17pc]{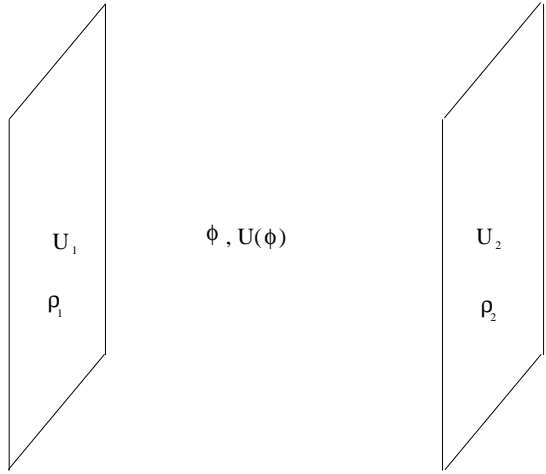}}
\caption{The five--dimensional setup consists of two boundary 
branes and a bulk scalar field with bulk potential $U$. Furthermore, 
the scalar field induce a brane tension on each branes with opposite 
signs.}
\end{figure}

At low energies, 
the action can be obtained using the moduli space approximation and we
found \cite{Brax} that the theory can be described by the following action 
(written here in the Einstein frame):
\begin{eqnarray}
S_{\rm EF} &=& \frac{1}{16\pi G} 
\int d^4x \sqrt{-g}\left[ {\cal R} -  \frac{12\alpha^2}{1+2\alpha^2}
\frac{(\partial Q)^2}{Q^2} - \frac{6}{2\alpha^2 + 1}(\partial
R)^2\right] \nonumber \\
&-& \int d^4 x\sqrt{-g} (V_{\rm eff}(Q,R)+W_{\rm eff}(Q,R)) \nonumber \\
&-& S_m^{(1)}(\Psi_1,A^2(Q,R)g_{\mu\nu}) 
- S_m^{(2)}(\Psi_2,B^2(Q,R)g_{\mu\nu}).
\end{eqnarray}
In this expression, $Q$ and $R$ are two scalar degrees of freedom
with potential energies $V_{\rm eff}$ and $W_{\rm eff}$. 
$S_m^{(i)}$ denote the matter action for each matter types.

The cosmological evolution equations can found to be:
\begin{equation}\label{Friedmann}
H^2 = \frac{8 \pi G}{3} \left(\rho_1 + \rho_2 + V_{\rm eff} 
+ W_{\rm eff} \right) + \frac{2\alpha^2}{1 + 2\alpha^2} \dot\phi^2
+ \frac{1}{1+2\alpha^2} \dot R^2.
\end{equation}
where we have defined $Q=\exp \phi$. 
The field equations for $R$ and $\phi$ read
\begin{eqnarray}
\frac{6}{1+2\alpha^2}\left[\ddot R + 3 H \dot R \right] 
&=& - 8 \pi G \left[ \frac{\partial V_{\rm eff}}{\partial R} + 
\frac{\partial W_{\rm eff}}{\partial R} \right. \nonumber \\
&+& \left. 2\alpha_R^{(1)} (\rho_1 - 3p_1) 
+ 2\alpha_R^{(2)} (\rho_2 - 3p_2) \right]
\end{eqnarray} 
\begin{eqnarray}
\ddot \phi + 3 H \dot \phi &=&
-8 \pi G \frac{1+2\alpha^2}{12 \alpha^2} [ 
\frac{\partial V_{\rm eff}}{\partial \phi} \nonumber \\ 
&+& \frac{\partial W_{\rm eff}}{\partial \phi} + 
2\alpha_\phi^{(1)} (\rho_1 - 3p_1) + 
2\alpha_\phi^{(2)} (\rho_2 - 3p_2) ].\label{Qcos}
\end{eqnarray}
Finally, the coupling parameters are given by
\begin{eqnarray}
\alpha_\phi^{(1)} &=& -\frac{2\alpha^2}{1+2\alpha^2}, \hspace{0.5cm}
\alpha_\phi^{(2)} = -\frac{2\alpha^2}{1+2\alpha^2} \label{coupling1}, \\
\alpha_R^{(1)} &=& \frac{\tanh R}{1+2\alpha^2}, \hspace{0.5cm}
\alpha_R^{(2)} = \frac{(\tanh R)^{-1}}{1+2\alpha^2} \label{coupling2}.
\end{eqnarray}

There is an obervational constraint for models described by an action 
like the one above \cite{Damour1}, namely the parameter $\theta$, 
defined by 
\begin{equation}
\theta= \frac{4}{3}\frac{\alpha^2}{1+2\alpha^2} 
+ \frac{\tanh^2 R}{6(1+2\alpha^2)}
\end{equation}
has to be less than $10^{-3}$. This implies, that 
\begin{equation}\label{constraint}
\alpha < 10^{-2},\ R < 0.2.
\end{equation}

The question is, therefore, {\it whether} $R$ {\it is driven towards 
small values during the cosmological evolution}. Furthermore, 
even if such an attractor exists, it is not clear if the 
attractor is efficient enough. To find an answer to these questions, 
we have numerically solved the equations of motions for different 
cases, which we will describe in the next section.

\section{Cosmological Evolution}
There are different cases to study. For example, 
there could be no matter on the negative tension brane or  
no potentials for the fields. The equations 
of motion of $R$ and $\phi$ show that their evolution depends on the 
matter content on the branes. We will assume both radiation and matter
on the positive tension brane and will follow the evolution of the
fields during the radiation and matter dominated epoch for the cases 
without and with matter on the negative tension brane. 

\subsection{No matter on the negative tension brane}
We assume no potential energy for the fields $R$ and $\phi$. 
In this case, the evolution of $R$ and $\phi$ are shown in 
fig. 2 and fig. 3.

\begin{figure}
\centerline{\includegraphics[width=17pc]{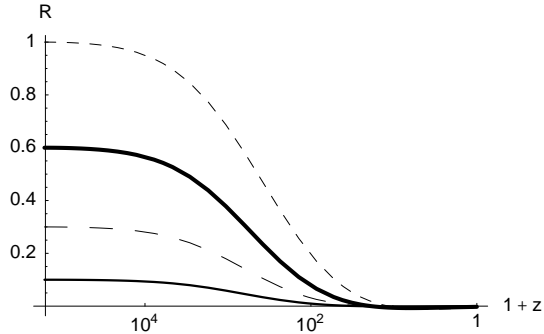}}
\caption{Evolution of the field $R$ during radiation and 
matter dominated epochs for the case of no matter on the negative 
tension brane.} 
\end{figure}

\begin{figure}
\centerline{\includegraphics[width=17pc]{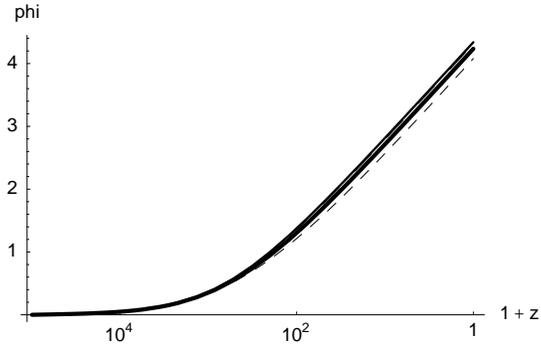}}
\caption{Evolution of the field $\phi$ during radiation and 
matter dominated epochs for the case of no matter on the negative 
tension brane.} 
\end{figure}

During the radiation dominated epoch, the fields $\phi$ and $R$ do 
not evolve strongly. This is because any time--evolution is 
sourced by the trace of the energy--momentum tensor, which vanishes  
if radiation dominates the expansion. During matter 
domination, one can clearly see that $R$ is driven towards 
small values for a wide range of initial conditions. 
Also, the evolution of $\phi$ does not strongly depend on the initial 
conditions. Thus, we conclude that there is an attractor 
for the field $R$, when there is no matter on the second brane. 

\subsection{Including matter on the negative tension brane}
We still assume no potential energy for the fields $R$ and $\phi$. 
The evolution of $R$ and $\phi$ are shown in figs. 4 and 5.

\begin{figure}
\centerline{\includegraphics[width=17pc]{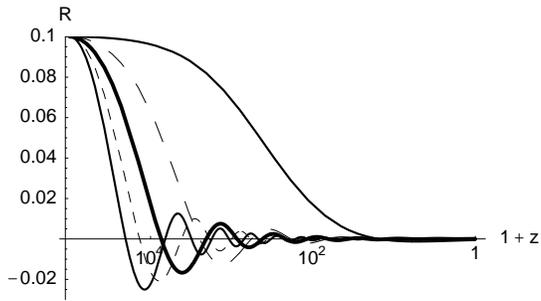}}
\caption{Evolution of the field $R$ during radiation and 
matter dominated epochs for the case with matter on the negative 
tension brane.} 
\end{figure}

\begin{figure}
\centerline{\includegraphics[width=17pc]{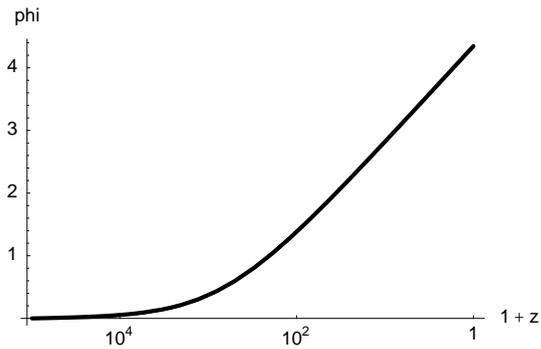}}
\caption{Evolution of the field $\phi$ during radiation and 
matter dominated epochs for the case with matter on the negative 
tension brane.} 
\end{figure}

We see, that in this case, the attractor is more effective. 
Thus, in both cases, {\it the field $R$ is driven towards 
small values.} In \cite{Brax} we have also included potentials for 
the fields, which might be necessary for the theory to be consistent. 
We refer to \cite{Brax} for more details. 

\section{Implications for varying constants}
In the Einstein frame, masses of particles vary, because
the two scalar fields $\phi$ and $R$ couple to the matter on both 
branes. Alternatively, it is possible to consider the frame in which 
the masses of particles living on the positive tension brane are 
constant; in that case, the gravitational constant is
time--varying as well as the masses of particles confined on the 
negative tension brane. These conclusions are generic for the kind of 
theories we consider. The constraint (\ref{constraint}) ensures that 
the time--variation is consistent with current observations. 
(There are constraints on the amplitude of $R$ and $\phi$ 
in the early universe from nucleosynthesis. We do not discuss this 
issue here.)

Is there any time variation for the electromagnetic 
fine structure constant? As explained in \cite{Brax}, this is not 
the case, as long as vector bosons are not coupled to the bulk scalar 
field (and therefore to $\phi$ and $R$). Thus, what is needed in order 
to explain the variation of the fine--structure constant is a 
coupling of the form 
\begin{equation}
S_m=-\frac{1}{4g^2}\int\sqrt{-g_B}f(R,\phi)g_B^{\mu\rho}
g_B^{\nu\eta}F_{\mu\rho}F_{\nu\eta}~~.
\end{equation}
Such a coupling appears naturally in heterotic M--theory, for 
example \cite{Witten}. In this theory the bulk scalar field 
has an geometrical interpretation: it is a measure of the 
deformation of a Calabi--Yau threefold. Generally, one would 
expect that the function $f$ is a function of the parameter 
$\alpha$ and that for $\alpha \rightarrow 0$ one has 
$f\rightarrow 1$, because for $\alpha = 0$ the bulk scalar 
field decouples from the dynamics \cite{Brax}. The attractor 
behaviour found in the last section implies, that any time--variation 
of the gauge--couplings at redshift between 0 and 5 is mainly due to 
the evolution of $\phi$. The amount of variation depends on the 
function $f$ above.

To conclude, the attractor behaviour we have identified might be
related to the stabilization of the moduli fields in brane world 
scenarios. When the moduli fields are not stabilised, an evolution 
of masses or the gravitational constant (depending on the frame) 
is generally predicted, whereas an evolution of the fine--structure constant 
is not generically predicted. Only if vector bosons are directly 
coupled to the bulk scalar field one would expect coupling constants to 
vary. This is the case in models based on string theory and therefore
one would necessary predict a time--variation of coupling constants as 
long as the moduli fields are not stabilized at a minimum of some potential. 

\vspace{0.1cm}\noindent {\bf Acknowledgements:} C.v.d.B. is grateful
to Malcolm Fairbairn for useful discussions on varying constants in 
M--theory. 
This work was supported by PPARC (C.v.d.B., A.-C.D. and C.S.R.), 
a CNRS--Royal Society exchange grant for collaborative research 
and the European network (RTN), HPRN--CT--200-00148 and 
PRN--CT--2000--00148.

\end{document}